\documentclass[twocolumn,secnumarabic,amssymb,nobibnotes,aps,showpacs,nofootinbib,noeprint,superscriptaddress]{revtex4-1}
\usepackage[english]{babel}
\usepackage{graphicx}

\usepackage{amsmath}
\usepackage{amsfonts}
\usepackage[usenames,dvipsnames]{color}

\usepackage{hyperref}
\usepackage{rotating}
\usepackage{booktabs} 
\usepackage{slashed}
\usepackage{transparent}

\usepackage{color}
\usepackage{soul}

\newcommand{\dd}{\; \mathrm{d}}

\DeclareMathAlphabet{\mathbb}{U}{bbold}{m}{n}

\usepackage{ulem}

\usepackage{hyperref}
\usepackage{nameref}

\begin{document}
\title{Screening lengths in ionic fluids}

\author{Fabian Coupette}
\affiliation{Institute of Physics, University of Freiburg, Hermann-Herder-Stra{\ss}e 3, 79104 Freiburg, Germany}

\author{Alpha A. Lee}
\affiliation{Cavendish Laboratory, JJ Thomson Ave., Cambridge CB3 0HE, United Kingdom }

\author{Andreas H\"{a}rtel}
\email{andreas.haertel@physik.uni-freiburg.de}
\affiliation{Institute of Physics, University of Freiburg, Hermann-Herder-Stra{\ss}e 3, 79104 Freiburg, Germany}

\date[]{published in: Physical Review Letters \textbf{121}, 075501 (2018), DOI: \href{https://doi.org/10.1103/PhysRevLett.121.075501}{10.1103/PhysRevLett.121.075501}}

\begin{abstract}
The decay of correlations in ionic fluids is a classical problem in soft matter physics that underpins 
applications ranging from controlling colloidal self-assembly to batteries and supercapacitors. 
The conventional wisdom, based on analyzing a solvent-free electrolyte model, suggests that all 
correlation functions between species decay with a common decay length in the asymptotic far 
field limit. Nonetheless a solvent is present in many electrolyte systems. 
We show using an analytical theory and molecular dynamics simulations that multiple decay lengths can coexist 
in the asymptotic limit as well as at intermediate distances once a hard sphere solvent is considered. 
Our analysis 
provides an explanation for the recently observed discontinuous change in the structural force 
across a thin film of ionic liquid-solvent mixtures as the composition is varied, as well as 
reframes recent debates in the literature about the screening length in concentrated electrolytes. 
\end{abstract}

\maketitle

The study of ionic fluids and electrolytes has received significant interest in recent times due 
to its central relevance to a plethora of technological applications, ranging from controlling colloidal 
self-assembly \cite{evans_book_1999} to supercapacitors and batteries \cite{fedorov_cr114_2014}. 
The challenge deals with the rich physics that arise from competing long-ranged Coulomb interactions and 
the steric repulsion of particles. 
The arrangement of ions in bulk and near interfaces governs properties such as 
capacitance \cite{bozym_jpcl6_2015,limmer_prl115_2015,uralcan_jpcl7_2016} and 
effective forces between colloids \cite{zhang_nature542_2017}; thus a physics understanding 
of how ion-ion correlations decay and how electric fields are screened is central to designing fit 
for purpose electrolytes. 

The decay of correlations in ionic fluids is a classical problem in 
soft matter and liquid state physics \cite{attard_inbook_2007,levin_rpp65_2002}. 
According to the conventional wisdom, all correlation functions in a simple fluid mixture 
where particles interact via short-ranged and Coulomb interactions 
decay asymptotically in the same form, 
\textit{i.e.}, $e^{-r/\lambda} \cos(\omega r - \tau)/r$, and, crucially, 
the decay length $\lambda$ -- synonymously the screening length -- and oscillation 
frequency $\omega$ are the same 
for all correlation functions \cite{evans_jcp100_1994}. 
This common decay has been explicitly verified 
for the restricted primitive model (RPM), a simple binary solvent-free electrolyte model 
that is paradigmatic in electrolyte physics -- it has been shown that the cation-cation, 
cation-anion and anion-anion correlation functions all decay with the same decay length 
and oscillation frequency \cite{attard_pre48_1993,leotedecarvalho_mp83_1994}, which has also 
been used for the interpretation of experiments \cite{zeng_langmuir28_2012,gebbie_pnas112_2015,smith_prl118_2017}. 
However, in technological applications, ions are usually mixed with a solvent in order to 
enhance conductivity and reduce viscosity 
\cite{mcewen_jes146_1999,zhu_science332_2011,yang_science341_2013}. This raises the important 
question of how the presence of solvents influences ion-ion correlations.

Recent surface force balance experiments show that the disjoining force between charged 
surfaces across ionic liquid-solvent mixtures decays in an oscillatory manner with an 
exponentially decaying envelope \cite{moazzami-gudarzi_prl117_2016,smith_prl118_2017,
schoen_bjnt9_2018}. However, as the ion concentration is increased, the oscillation frequency 
undergoes a steplike transition \cite{smith_prl118_2017} -- at low ion concentration, 
it is comparable to the size of the solvent molecule, whereas for concentrated 
electrolytes it is comparable to the size of an ion pair. This is qualitatively 
reminiscent of structural crossover in a binary mixture of ``big'' and ``small'' colloids 
\cite{grodon_jcp121_2004,baumgartl_prl98_2007,statt_jcp144_2016}. However, an ion-solvent 
mixture is evidently at least a three component system and a corresponding mechanism in 
electrolyte-solvent mixtures is, perhaps surprisingly, hitherto unknown. 

In this Letter, we demonstrate that the decay of correlation functions in a simple fluid mixture 
is not necessarily unique, \textit{i.e.}, there is no common asymptotic decay length and oscillation 
wavelength. By considering a hard sphere electrolyte in a hard sphere solvent -- one of the simplest 
possible extensions of the paradigmatic RPM model that includes the physics of 
electrolyte-solvent interactions -- we show theoretically that 
ion-ion correlations and ion-solvent correlations can have different asymptotic decay lengths 
and support this result using simulations. 
These decays are either density- or charge-driven and related to the 
length scales of steric and Coulombic interactions. 
While ion-solvent correlations are not affected by charge correlations, 
ion-ion correlations decay according to a superposition of both effects. 
However, asymptotic decay is determined by the slowest decaying contribution, 
which strongly varies with the system composition. Our theory explains the experimentally 
observed switch of the structural force as the crossover from density-driven to charge-driven decay 
\cite{smith_prl118_2017}. 
Moreover, it illustrates the importance of space-filling solvent, an often overlooked piece of 
physics in the theoretical modeling of electrolytes. 

To concretize ideas, we consider a hard sphere ion-solvent mixture (HISM) 
\cite{grimson_cpl86_1982,tang_jcp97_1992,boda_jcp112_2000,rotenberg_jpcm30_2018} throughout this 
Letter: ions and solvent are modeled as hard spheres of the same diameter $d$. The 
ions (solvent) with number density $\rho$ ($\rho_0$) carry point charges $Z_{\pm} = \pm e$ ($Z_{0} = 0$). 
The dielectric nature of the solvent is modeled by a homogeneous dielectric background with a 
relative permittivity $\varepsilon$. The pair interaction potential between two particles of 
species $\nu,\nu' \in \{+,-,0\}$ at separation $r$ is given by 
\begin{align}
v_{\nu\nu'}(r) &= 
\begin{cases}
\displaystyle{\infty } & r < d  \\
\displaystyle{k_{\rm B}T \lambda_{\rm B} \frac{Z_{\nu} Z_{\nu'}}{r}} & r \geq d , 
\end{cases}
\label{eq:pair-interaction-potential}
\end{align}
where $\lambda_{\rm B}=e^2/(4\pi\varepsilon_0\varepsilon k_{\rm B}T)$ denotes the Bjerrum 
length and $k_{\rm B}$ Boltzmann's constant.

\begin{figure}[t]
	\centering
	\includegraphics[width=8.5cm]{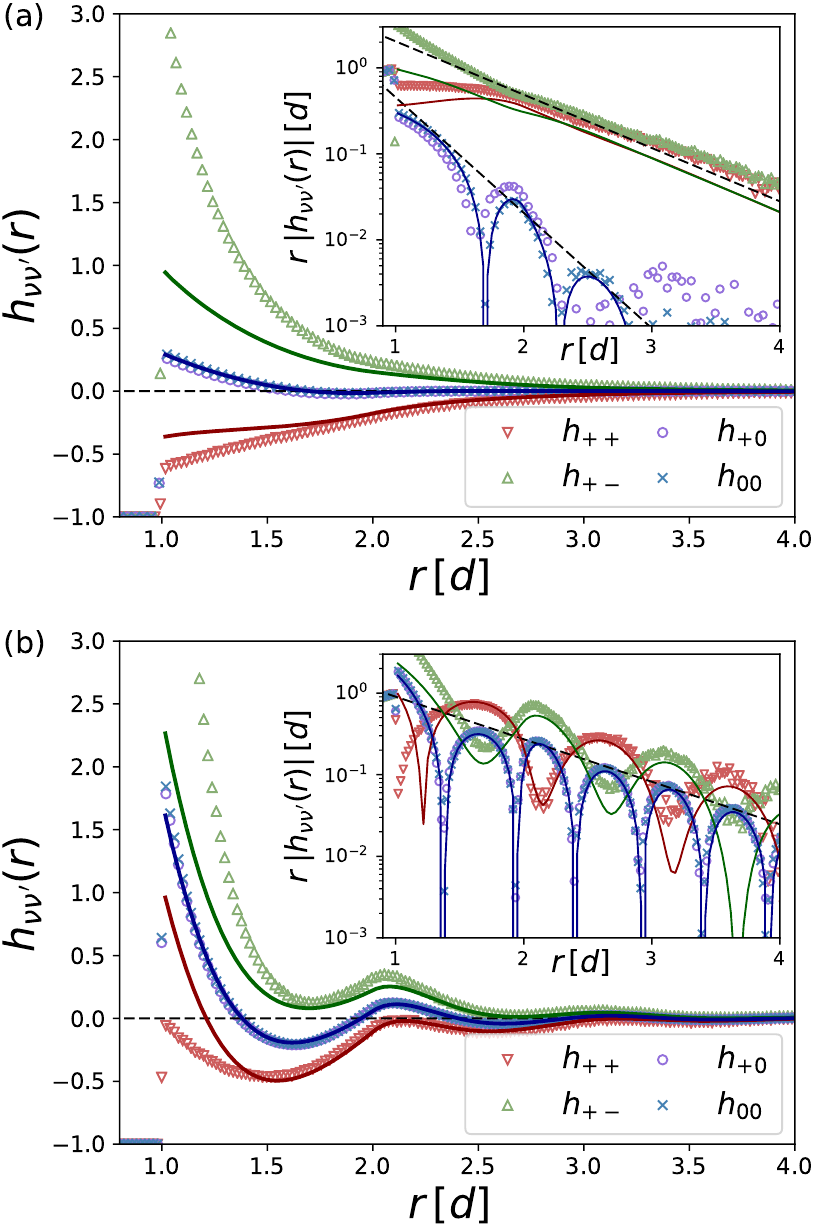}
	\caption{\label{fig:screening-length}
	Total pair-correlation functions $h_{\nu\nu'}(r)$ for ion concentration $\rho=1$ M 
	and concentration of neutral particles (a) $\rho_{0}=10$ M and (b) $\rho_{0}=40$ M. 
	Symbols represent data from MD simulations and lines from our theory. 
	For symmetry reasons, we only show data for the four given combinations of species. 
	The insets show the same data but plotted on semilogarithmic scale. Dashed lines represent 
	the predicted monotonic decay $\exp(-r/\lambda_{\nu\nu'})$ with screening length $\lambda_{\nu\nu'}$ 
	from theory. 
	}
\end{figure}

Figure~\ref{fig:screening-length} shows that the HISM model can have two distinguished coexisting 
screening lengths at finite range. We performed MD simulations of the HISM in an equilibrated bulk 
system using the ESPResSo package \cite{limbach_cpc174_2006,arnold_book_2013}. 
Hard particle interactions are modeled using a shifted and truncated purely repulsive 
Lennard-Jones potential $4\epsilon[(\sigma/r)^{12}-(\sigma/r)^{6}+c_{\rm shift}]$ with 
$\epsilon=10^4 \, k_{\rm B}T$ and $\sigma=2^{-1/6} \, d$. The simulations are performed in 
a cubic box of volume $V=L\times L\times L$ with periodic boundaries and $L=30 \, d$. 
We used $d=0.3$ nm and $\lambda_{\rm B}=0.7$ nm, which corresponds to $\varepsilon\approx 80$ 
and $T\approx 300$ K. At ionic concentration $\rho=1$ M and solvent concentration $\rho_0=10$ M, 
Fig.~\ref{fig:screening-length}(a) clearly shows two coexisting decay lengths with oscillatory 
and purely exponential decay, respectively, at intermediate separations. 
Figure~\ref{fig:screening-length}(b) shows that at a higher solvent concentration $\rho_0=40$ M, both 
ion-ion and ion-solvent correlations share the same intermediate decay length and oscillation wavelength. 
Our theory (see below) predicts that this finite range decay is the same as the asymptotic decay. 

To explain the origin of those coexisting decay lengths, we turn to a theoretical description 
of HISM based on the density functional theory (DFT) formalism \cite{hansen_book_2013}. 
Within DFT, the free energy is expressed as a functional of one-body densities \cite{hansen_book_2013}. 
For HISM, we can split the pair potential into hard core and electrostatic contributions, 
$v_{\nu \nu'} = v^{\mathrm{hs}}_{\nu \nu'}  +v^{\mathrm{es}}_{\nu \nu'}$. The difference 
between ideal gas free energy and the exact free energy can be partitioned into three 
components \cite{hansen_book_2013}, 
\begin{align}
F = F^{\rm hs} + F^{\rm es} + F^{\rm corr} , 
\label{eq:functional_expansion}
\end{align}
where $F^{\rm hs}$ is the hard sphere contribution, $F^{\rm es}$ the electrostatic contribution, 
and $F^{\rm corr}$ a correlation term that contains remaining contributions. 
The splitting in Eq.~(\ref{eq:functional_expansion}), although mathematically trivial, 
allows us to identify symmetries in the corresponding direct correlations $c_{\nu\nu'}^{\rm hs}$, 
$c_{\nu\nu'}^{\rm es}$, and $c_{\nu\nu'}^{\rm corr}$. The latter follow from a second functional 
derivative of the excess free energy with respect to the density, \textit{i.e.}, 
\begin{align}
c_{\nu\nu'}(r) 
&= - \frac{1}{k_{\rm B}T} 
\frac{\delta^2 F}{\delta \rho_{\nu}({\bf r_1}) \delta \rho_{\nu'}({\bf r_2)}} , 
\end{align} 
where the homogeneity of the bulk implies $r=|{\bf r_1}-{\bf r_2}|$. 
The hard sphere contribution depends only on the macroscopic packing fraction and the particle 
diameter $d$, and therefore, it scales equally with the number density of each component. 
From $F^{\rm es}$ given by \cite{hansen_book_2013} 
\begin{align}
F^{\rm es} &= \frac{1}{2} \sum_{\nu} \sum_{\nu'} 
\int\int \rho_\nu( {\bf r}) \rho_{\nu'}({\bf r'}) v^{\rm es}_{\nu \nu'}({\bf r},{\bf r'}) \dd {\bf r} \dd { \bf r'} , 
\end{align}
the electrostatic contribution follows with 
\begin{align}
c_{\nu\nu'}^{\rm es}(r) &= -\frac{v^{\rm es}_{\nu \nu'}(r)}{k_{\rm B}T} . 
\label{eq:es-c}
\end{align}
Hence, $c^{\rm es}:= c_{++}^{\rm es} = -c_{+-}^{\rm es}$. 
Finally, the correlation term underlies the fundamental symmetries of the system, \textit{i.e.}, 
positive and negative ions are structurally equivalent such that 
$c_{++}^{\rm corr} = c_{--}^{\rm corr}$, $c_{+-}^{\rm corr} = c_{-+}^{\rm corr}$, and 
$c_{+0}^{\rm corr} = c_{-0}^{\rm corr} = c_{0-}^{\rm corr} = c_{0+}^{\rm corr}$. 

The decomposition in Eq.~(\ref{eq:functional_expansion}) entails that the most general form of the 
direct correlation matrix $\mathcal{C}$ in the species basis $\{+,-,0\}$ for the HISM model is given by 
\begin{align}
\mathcal{C} &= 
\begin{pmatrix}
c^{\rm hs} + c^{\rm es} + c_{++}^{\rm corr} & c^{\rm hs} - c^{\rm es} + c_{+-}^{\rm corr} & c^{\rm hs} + c_{+0}^{\rm corr} \\
c^{\rm hs} - c^{\rm es} + c_{+-}^{\rm corr} & c^{\rm hs} + c^{\rm es} + c_{++}^{\rm corr} & c^{\rm hs} + c_{+0}^{\rm corr} \\
c^{\rm hs} + c_{+0}^{\rm corr} & c^{\rm hs} + c_{+0}^{\rm corr} & c^{\rm hs} + c_{00}^{\rm corr}
\end{pmatrix}. 
\label{eq:hism-c}
\end{align}
To proceed, we need to relate the direct correlation functions to the total correlation functions 
$h_{\nu\nu'}=(\mathcal{H})_{\nu\nu'}$, the observables in simulations and experiments. We use 
the Ornstein-Zernike relation in Fourier space 
\begin{align}
\hat{\mathcal{H}} &= \left( \mathbb{1}-\hat{\mathcal{C}} \varrho \right)^{-1} \hat{\mathcal{C}} , 
\label{eq:oz-fourier}
\end{align}
where we introduced a number density matrix $\varrho = \mathrm{diag}(\rho,\rho,\rho_0)$, 
and $\hat{f}$ denotes the Fourier transformation of a function $f$. 
Substituting Eq.~(\ref{eq:hism-c}) into (\ref{eq:oz-fourier}) yields an algebraic expression for 
the total correlation matrix $\hat{\mathcal{H}}$, the eigenvectors of which are given by 
${\bf w}_{\rm cc}=(1,-1,0)$, ${\bf w}_{\rm dd}^{+}$, and ${\bf w}_{\rm dd}^{-}$. 
The former is equal to one of 
the eigenvectors of the RPM and gives rise to the well established charge-charge correlation 
$h_{\rm cc}=h_{++}-h_{+-}$ as an eigenvalue \cite{hansen_book_2013}. The eigenvectors 
${\bf w}_{\rm dd}^{\pm}$ become stationary in the limit of vanishing $c_{\nu\nu'}^{\rm corr}$ 
with $(1,1,1)$ and $(-1,-1,2)$; the first of them gives rise to a density-density correlation, 
while the second corresponds to an ion-solvent correlation that has a vanishing eigenvalue. 

In particular, the resulting total charge-charge correlation function reads 
\begin{align}
\hat{h}_{\rm cc} &= 
\frac{\hat{c}^{\rm corr}_{ \rm cc} + 2 \hat{c}^{\rm es} }
{1 - \rho ( \hat{c}^{\rm corr}_{ \rm cc} + 2 \hat{c}^{\rm es}) } . 
\label{eq:hcc}
\end{align}  
Transforming it back into real space yields the formal solution 
\begin{align}
h_{\rm cc}(r) 
&= \frac{1}{2 \pi^2 r} \int_0^\infty k \sin(k r)  
	\frac{\hat{c}^{\rm corr}_{ \rm cc} + 2 \hat{c}^{\rm es} }
			{1 - \rho ( \hat{c}^{\rm corr}_{ \rm cc} + 2  \hat{c}^{\rm es})}
	\dd k  \nonumber \\
&= \frac{1}{2 \pi r}\sum_{q \in Q_{\rm cc}}  
	\Re \left[ \mathrm{Res}_q \left\{   
		\frac{ (\hat{c}^{\rm corr}_{ \rm cc} + 2 \hat{c}^{\rm es}) q \exp(i q r) }
				{1 - \rho ( \hat{c}^{\rm corr}_{ \rm cc} + 2 \hat{c}^{\rm es})   } 	
	\right\} \right] , 
\label{eq:residue}
\end{align}
where $Q_{\rm cc}$ contains the roots of 
\begin{align}
1 - \rho ( \hat{c}^{\rm corr}_{ \rm cc} + 2  \hat{c}^{\rm es}) &= 0 
\label{eq:root}
\end{align}
with positive imaginary part. 
The second equality in Eq.~(\ref{eq:residue}) makes use of the residue theorem and does, therefore, 
only hold without further analysis if Eq.~(\ref{eq:root}) does not have any purely real solutions 
and the elements of $Q_{\rm cc}$ are isolated points in the upper complex half plane 
(we refer to Refs.~\cite{kjellander_cpl200_1992,attard_pre48_1993,leotedecarvalho_mp83_1994,evans_jcp100_1994} 
for similar derivations). 
The eigenvalues to ${\bf w}^\pm_{\rm dd}$ share a common denominator, \textit{i.e.}, they are 
a set $Q_{\rm dd}$ of singularities corresponding to the roots 
with positive imaginary part of the generic equation 
\begin{align}
1 - \rho (2\hat{c}^{\rm hs} + \hat{c}_{\rm dd}^{\rm corr}) &~ \notag \\
- \rho_0 [ 2\rho(\hat{c}_{+0}^{\rm corr})^2 + \hat{c}_{00}^{\rm corr} (1-\rho\hat{c}_{\rm dd}^{\rm corr}) &~ \notag \\
+ \hat{c}^{\rm hs}(1 - \rho (\hat{c}_{\rm dd}^{\rm corr}-4\hat{c}_{+0}^{\rm corr}+2\hat{c}_{00}^{\rm corr}))] &= 0 , 
\label{eq:root2}
\end{align}
where $c_{\rm dd}^{\rm corr}:= c_{++}^{\rm corr}+c_{+-}^{\rm corr}$. 
Note that Eq.~\ref{eq:root2} is independent of $c^{\rm es}$. 

The dominant contribution to a total correlation function $h_{\nu\nu'}$ in the asymptotic long-range limit 
$r\to\infty$ is determined by the (leading) pole 
$\bar{q}_{\nu \nu'}=\Re[\bar{q}_{\nu \nu'}] +i \Im[\bar{q}_{\nu \nu'}] \in Q_{\nu\nu'}$ 
with the smallest imaginary part \cite{fisher_jcp50_1969}. 
It is convenient to introduce the decay length $\lambda_{\nu \nu'}=1/\Im[\bar{q}_{\nu \nu'}]$ 
and decay oscillation frequency $\omega_{\nu \nu'}=\Re[\bar{q}_{\nu \nu'}]$. 
This pole causes the asymptotic decay \cite{fisher_jcp50_1969,henderson_jcp97_1992,evans_jcp100_1994} 
\begin{align}
h_{\nu\nu'}(r \rightarrow \infty) \propto \frac{\exp(- r/\lambda_{\nu\nu'} ) \cos(\omega_{\nu\nu'} r-\tau_{\nu\nu'})}{r} , 
\label{eq:fit}
\end{align}
where $\tau_{\nu\nu'}$ is a phase shift. 
The pole, however, could be suppressed on intermediate 
length scales by a small amplitude such that its contribution would become neglectable. 
If there are two poles with decay lengths $\lambda_1>\lambda_2$ but amplitudes $A_1<A_2$, 
pole 2 will dominate until $r\gtrsim\log(A_1/A_2)[\lambda_1^{-1}-\lambda_2^{-1}]^{-1}$, 
which is a long length scale if $A_1 \ll A_2$. 

Importantly, two competing decay lengths arise from the solutions to Eqs.~(\ref{eq:root}) and (\ref{eq:root2}). 
Switching back into the species basis yields the central result of this Letter,  
\begin{align}
\lambda_{++} = 
\lambda_{+-} &= \max[\lambda_{\rm cc}, \lambda_{\rm dd}] , \label{eq:lambda-cc} \\
\lambda_{0+} = \lambda_{0-} = \lambda_{00} &= \lambda_{\rm dd} . \label{eq:lambda-dd}
\end{align}
The charge-charge correlation does not affect solvent correlations, because ${\bf w}_{\rm cc}\perp(0,0,1)$. 
Notice that we only made use of the fundamental symmetries in HISM. In other words, 
in general, it is not true that all species correlations decay with the same decay length. 
Correlations involving solvent particles decay on a length scale $\lambda_{\rm dd}$ different 
from the charge-charge correlation length scale $\lambda_{\rm cc}$. 
If $\lambda_{\rm cc} > \lambda_{\rm dd}$, two distinct length scales coexist, 
as we have shown for intermediate ranges in Fig.~\ref{fig:screening-length}(a). 
The same applies for the corresponding oscillation frequencies $\omega_{\rm cc}$ and $\omega_{\rm dd}$. 
Crucially, this implies that while the dominant decay length continuously changes, 
the oscillation wavelength of ion-ion correlations can rapidly shift. 

\begin{figure}[t]
	\centering
	\includegraphics[width=8.5cm]{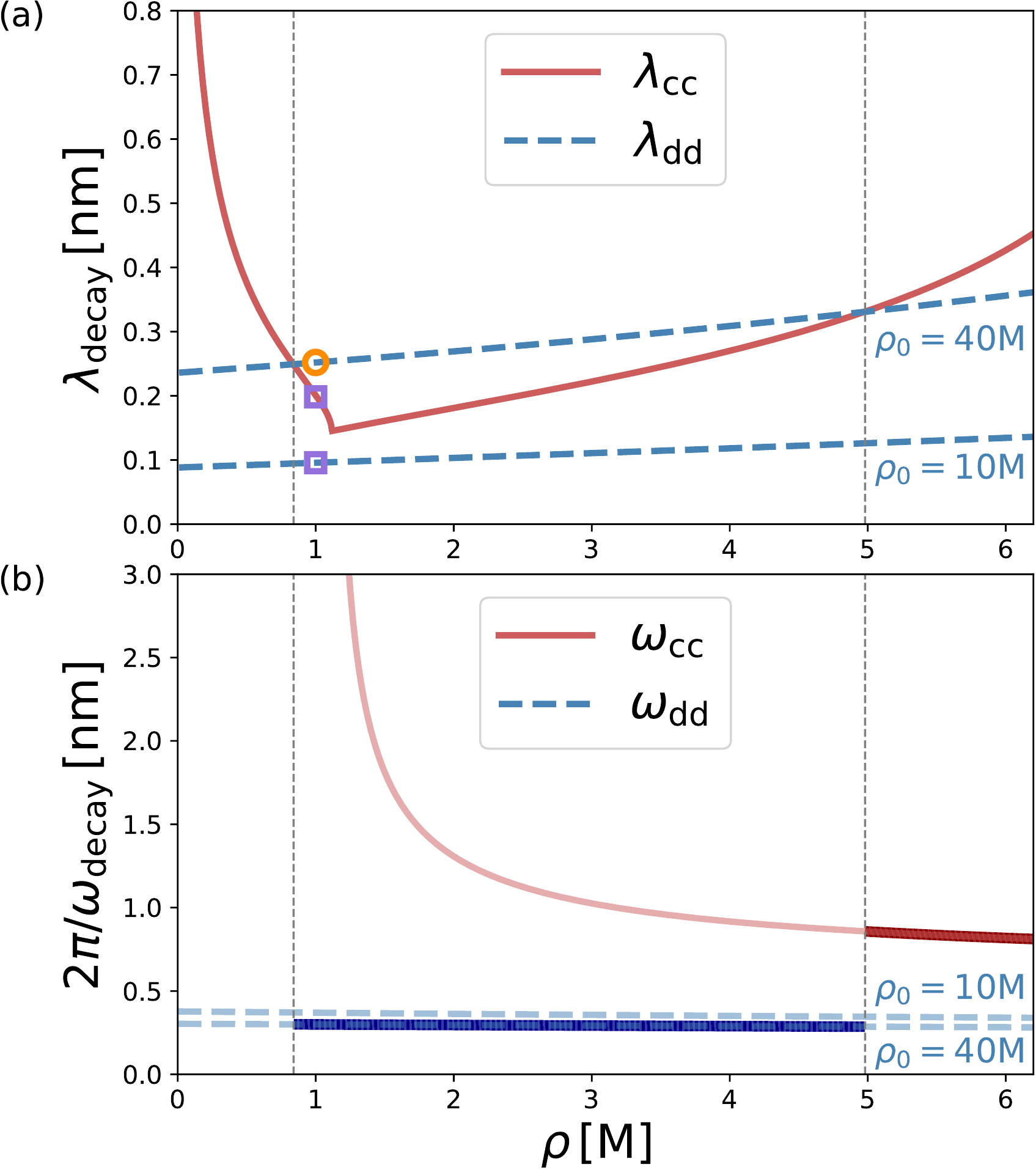}
	\caption{\label{fig:screening-regimes}
	Theoretical prediction for 
	(a) decay length and (b) inverse oscillation frequency of the leading charge and density pole, 
	respectively, shown against the ion concentration $\rho$, for $\rho_0=10$ M and $\rho_0=40$ M 
	with $d=0.3$ nm and $\lambda_{\rm B}=0.7$ nm. 
	Symbols in (a) mark the decay lengths that correspond to the data in 
	Figs.~\ref{fig:screening-length}(a) ($\square$) and \ref{fig:screening-length}(b) ($\bigcirc$). 
	Vertical dashed lines mark points where the asymptotically leading pole changes 
	from charge to density and \textit{vice versa} for $\rho_0=40$ M; leading 
	inverse oscillation frequencies for $\rho_0=40$ M are highlighted with bold lines. 
	}
\end{figure}

To illustrate this effect, we proceed by specifying the functional $F$ in our theoretical framework: 
we use the White Bear mark II functional for the hard-sphere contribution \cite{hansen-goos_jpcm18_2006} and 
Eq.~(\ref{eq:es-c}) with $v_{\nu\nu'}^{\rm es}(r)=0$ for $r<d$ for the electrostatic term. 
By setting $c^{\rm corr}\equiv 0$ we obtain analytical correlation functions that are sufficient to 
illustrate the mechanism of the wavelength switch; for the observed systems deviations due to this 
approximation mainly occur at particle contact, as shown in Fig.~\ref{fig:screening-length}.

Figure~\ref{fig:screening-regimes} shows quantitative predictions of our theory for the decay lengths 
and the oscillation wavelengths in HISM. The density-induced correlation length, $\lambda_{\rm dd}$, 
is a monotonic function of the macroscopic volume fraction because steric correlations are enhanced 
as the system becomes denser. However, the charge-induced correlation length, $\lambda_{\rm cc}$, 
is a nonmonotonic function of the ion density but independent of the solvent density. 
Further, this is the length scale of the decay of the effective electrostatic potential that 
an ion generates.  
$\lambda_{\rm cc}$ decreases for an increasing ion density in a dilute electrolyte 
because ions are surrounded by counterions and this arrangement progressively screens the 
electric field that an ion generates. However, past a threshold ion concentration, 
ion-ion correlations lead to a counterion solvation shell that overcompensates the ionic charge, 
which causes a second solvation shell to solvate the counterions, triggering an oscillatory 
decay \cite{attard_inbook_2007}. In this regime, increasing the ion concentration amplifies 
ion-ion correlations; thus the screening length grows. The situation, 
when the charge pole that determines $\lambda_{\rm cc}$ changes from purely imaginary 
to complex, \textit{i.e.}, the decay changes from monotonic to oscillatory, is called a Kirkwood 
transition \cite{kirkwood_jcp7_1939}, and here it coincides with the change between decreasing 
and increasing screening length. 
When $\lambda_{\rm cc} > \lambda_{\rm dd}$, which is the case for a large region of RPM's 
parameter space, the ion-ion correlations decay with a decay length that is the electrostatic screening 
length but different from the ion-solvent and solvent-solvent correlations decay 
(Fig.~\ref{fig:screening-regimes}a). For a high solvent concentration, however, we find a regime 
$\lambda_{\rm dd} >\lambda_{\rm cc}$ where all species correlations decay with one common decay length 
$\lambda_{\rm dd}$ [see also Fig.~\ref{fig:screening-length}(b)] but different from the charge-charge 
decay length. Thus, the electrostatic screening length must be distinguished from the decay length of 
species correlation functions that is typically observed in experiments.

Although the ion-ion decay length switches continuously from one pole to another in 
Fig.~\ref{fig:screening-regimes}(a), Fig.~\ref{fig:screening-regimes}(b) shows that the 
corresponding oscillation frequency exhibits a discontinuous jump that occurs when the 
two leading poles have equal imaginary but different real parts. This jump is precisely 
the effect observed in experimental studies of the surface force across ion-solvent mixtures 
\cite{smith_prl118_2017} -- the oscillation wavelength switches abruptly. In the experiment, 
ions and solvent molecules are approximately of the same size, and the oscillation wavelength 
jumps from $d$ to $2d$, which agrees squarely with the prediction in 
Fig.~\ref{fig:screening-regimes}(b) (see the Supplemental Material \cite{SM} for a detailed comparison). 
Note that the position of this discontinuous jump in the oscillation wavelength is different 
from the onset of charge oscillations at the Kirkwood transition when the real part of the charge 
pole first takes a finite nonvanishing value \cite{parrinello_rnc2_1979,stell_prl37_1976}. 
Furthermore, the increase of the decay length in Fig.~\ref{fig:screening-regimes}(a) accurately 
describes the decay of the structural force in experiments \cite{smith_prl118_2017}. 
However, the experiments show an additional much longer decay length at large separations, 
which is neither predicted in our theory and other recent theoretical studies of 
underscreening \cite{lee_prl119_2017,rotenberg_jpcm30_2018} nor observed in our 
simulations on HISM (see Fig.~S2 in \cite{SM}). 
This long-ranged decay length might arise from a set of additional poles induced by a mechanism that 
is not contained in the simplified HISM model. For instance, dipolar solvent-solvent 
interactions, as present in water, could lead to an additional long decay length. Since this 
long-ranged decay is experimentally only observed at long distances, the corresponding leading 
pole should have a small amplitude and, therefore, could be suppressed at intermediate 
distances (see Fig.~\ref{fig:screening-length}).

\begin{figure}[t]
	\centering
	\includegraphics[width=8.0cm]{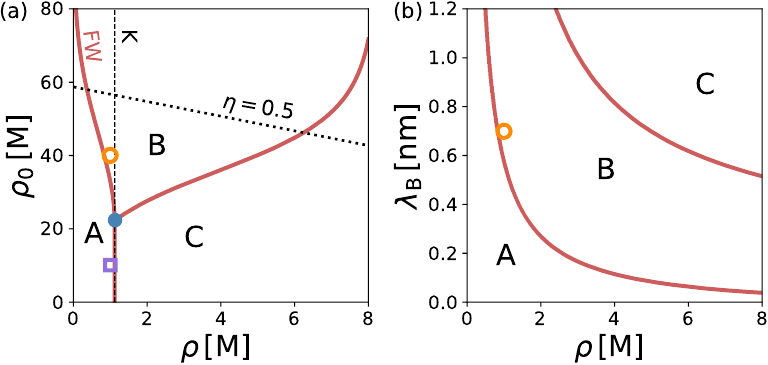}
	\caption{\label{fig:phase-diag}
	(a) Different regimes of decay in the $\rho$ -- $\rho_0$ plane of ion and neutral particle concentrations: 
	(A) purely exponential, charge-dominated, (B) oscillatory, exponentially damped, density-dominated, 
	and (C) oscillatory, exponentially damped, charge-dominated. The vertical dashed line represents the 
	Kirkwood line (K) \cite{kirkwood_jcp7_1939,leotedecarvalho_mp83_1994}. The right boundary of region A represents 
	the Fisher-Widom line (FW) \cite{fisher_jcp50_1969,evans_mp80_1993}. The tilted dashed line marks a 
	total volume fraction $\eta=0.5$. Symbols are explained in Fig.~\ref{fig:screening-regimes}. 
	(b) The same diagram in the $\rho$ -- $\lambda_{\rm B}$ plane for $\rho_0=40$ M. }
\end{figure}

The three different regimes of asymptotic decay in HISM -- purely exponential and charge-dominated 
decay (A), oscillatory exponentially damped and density-dominated decay (B), and oscillatory 
exponentially damped and charge-dominated decay (C) -- are summarized in Fig.~\ref{fig:phase-diag}. 
While ion-ion correlations in regions A and C are dominated by the charge pole, ion-ion correlations 
couple to the solvent in region B. This region appears at high solvent concentrations between A 
and C such that the Fisher-Widom line \cite{fisher_jcp50_1969,evans_mp80_1993} of the ions shifts towards lower ion 
concentrations (away from the Kirkwood line \cite{kirkwood_jcp7_1939,leotedecarvalho_mp83_1994}). 
A second branch separates regions A and C at which the frequency jumps from $\omega_{\rm dd}$ to $\omega_{\rm cc}$. 

Our conclusions are derived by assuming symmetry between positive and negative ions in Eq.~(\ref{eq:hism-c}). 
If this symmetry is broken by different ion sizes, all correlation functions couple and share the same 
set of poles; thus they all decay asymptotically in the same form. 
However, at intermediate range, simulations of asymmetric ions still exhibit the same coexistence of decay 
lengths and oscillation frequencies as shown here for the symmetric case \cite{SM}. 
Consequently, ion size asymmetry can be considered as a perturbation to the 
symmetric HISM model so that its predictions are still valid for decay lengths in asymmetric 
systems at (experimentally relevant) intermediate distances.

In summary, we demonstrated the possible coexistence of two asymptotic decay lengths for hard sphere ions 
in a hard sphere solvent. Our theory explains recent experimental findings concerning a jump of the wavelength 
of the structural force in ionic fluids \cite{smith_prl118_2017}, and it sheds new light on the screening in dense 
electrolytes and the fitting of structural forces \cite{schoen_bjnt9_2018}. 
Our results are important for the interpretation of measurements and effective interactions 
\cite{gottwald_prl92_2004,leger_jcp123_2005,denton_pre96_2017,schoen_bjnt9_2018}, 
because they show that species correlation functions can be superpositions of charge contributions 
and density contributions of the same order of magnitude. A fit using the asymptotic form (\ref{eq:fit}) 
hence cannot be expected to be accurate on intermediate length scales. 
Furthermore, the transition from monotonic to oscillatory decay underpins wetting phenomena 
\cite{chernov_prl60_1988,henderson_pre50_1994}. The existence of multiple coexisting species-dependent 
decay lengths implies that addressable wetting could be achieved. 
Tuning the asymptotic correlations may also be used to control colloidal dispersions,
for instance to prevent aggregation \cite{zhang_nature542_2017} and to 
switch effective potentials by tuning the salt concentration \cite{li_pnas114_2017}. 
It might be promising to construct complex interactions to achieve a rich crossover structure, 
for instance in complex plasmas \cite{morfill_rmp81_2009}, 
colloid-polymer mixtures \cite{brader_pre63_2001}, 
and colloidal fluids \cite{archer_jcp126_2007}.


\begin{acknowledgments}
The authors would like to thank M.~Oettel, R.~Kjellander, and R.~Evans 
for insightful discussions. 
A.~A.~L. acknowledges the support of the Winton Programme for the Physics of Sustainability. 
\end{acknowledgments}


\begin{thebibliography}{53}%
\makeatletter
\providecommand \@ifxundefined [1]{%
 \@ifx{#1\undefined}
}%
\providecommand \@ifnum [1]{%
 \ifnum #1\expandafter \@firstoftwo
 \else \expandafter \@secondoftwo
 \fi
}%
\providecommand \@ifx [1]{%
 \ifx #1\expandafter \@firstoftwo
 \else \expandafter \@secondoftwo
 \fi
}%
\providecommand \natexlab [1]{#1}%
\providecommand \enquote  [1]{``#1''}%
\providecommand \bibnamefont  [1]{#1}%
\providecommand \bibfnamefont [1]{#1}%
\providecommand \citenamefont [1]{#1}%
\providecommand \href@noop [0]{\@secondoftwo}%
\providecommand \href [0]{\begingroup \@sanitize@url \@href}%
\providecommand \@href[1]{\@@startlink{#1}\@@href}%
\providecommand \@@href[1]{\endgroup#1\@@endlink}%
\providecommand \@sanitize@url [0]{\catcode `\\12\catcode `\$12\catcode
  `\&12\catcode `\#12\catcode `\^12\catcode `\_12\catcode `\%12\relax}%
\providecommand \@@startlink[1]{}%
\providecommand \@@endlink[0]{}%
\providecommand \url  [0]{\begingroup\@sanitize@url \@url }%
\providecommand \@url [1]{\endgroup\@href {#1}{\urlprefix }}%
\providecommand \urlprefix  [0]{URL }%
\providecommand \Eprint [0]{\href }%
\providecommand \doibase [0]{http://dx.doi.org/}%
\providecommand \selectlanguage [0]{\@gobble}%
\providecommand \bibinfo  [0]{\@secondoftwo}%
\providecommand \bibfield  [0]{\@secondoftwo}%
\providecommand \translation [1]{[#1]}%
\providecommand \BibitemOpen [0]{}%
\providecommand \bibitemStop [0]{}%
\providecommand \bibitemNoStop [0]{.\EOS\space}%
\providecommand \EOS [0]{\spacefactor3000\relax}%
\providecommand \BibitemShut  [1]{\csname bibitem#1\endcsname}%
\let\auto@bib@innerbib\@empty
\bibitem [{\citenamefont {Evans}\ and\ \citenamefont
  {Wennerstr{\"o}m}(1999)}]{evans_book_1999}%
  \BibitemOpen
  \bibfield  {author} {\bibinfo {author} {\bibfnamefont {D.~F.}\ \bibnamefont
  {Evans}}\ and\ \bibinfo {author} {\bibfnamefont {H.}~\bibnamefont
  {Wennerstr{\"o}m}},\ }\href@noop {} {\emph {\bibinfo {title} {The colloidal
  domain}}}\ (\bibinfo  {publisher} {Wiley-Blackwell},\ \bibinfo {address} {New
  York},\ \bibinfo {year} {1999})\BibitemShut {NoStop}%
\bibitem [{\citenamefont {Fedorov}\ and\ \citenamefont
  {Kornyshev}(2014)}]{fedorov_cr114_2014}%
  \BibitemOpen
  \bibfield  {author} {\bibinfo {author} {\bibfnamefont {M.~V.}\ \bibnamefont
  {Fedorov}}\ and\ \bibinfo {author} {\bibfnamefont {A.~A.}\ \bibnamefont
  {Kornyshev}},\ }\href {\doibase 10.1021/cr400374x} {\bibfield  {journal}
  {\bibinfo  {journal} {Chem. Rev.}\ }\textbf {\bibinfo {volume} {114}},\
  \bibinfo {pages} {2978} (\bibinfo {year} {2014})}\BibitemShut {NoStop}%
\bibitem [{\citenamefont {Bozym}\ \emph {et~al.}(2015)\citenamefont {Bozym},
  \citenamefont {Uralcan}, \citenamefont {Limmer}, \citenamefont {Pope},
  \citenamefont {Szamreta}, \citenamefont {Debenedetti},\ and\ \citenamefont
  {Aksay}}]{bozym_jpcl6_2015}%
  \BibitemOpen
  \bibfield  {author} {\bibinfo {author} {\bibfnamefont {D.~J.}\ \bibnamefont
  {Bozym}}, \bibinfo {author} {\bibfnamefont {B.}~\bibnamefont {Uralcan}},
  \bibinfo {author} {\bibfnamefont {D.~T.}\ \bibnamefont {Limmer}}, \bibinfo
  {author} {\bibfnamefont {M.~A.}\ \bibnamefont {Pope}}, \bibinfo {author}
  {\bibfnamefont {N.~J.}\ \bibnamefont {Szamreta}}, \bibinfo {author}
  {\bibfnamefont {P.~G.}\ \bibnamefont {Debenedetti}}, \ and\ \bibinfo {author}
  {\bibfnamefont {I.~A.}\ \bibnamefont {Aksay}},\ }\href {\doibase
  10.1021/acs.jpclett.5b00899} {\bibfield  {journal} {\bibinfo  {journal} {J.
  Phys. Chem. Lett.}\ }\textbf {\bibinfo {volume} {6}},\ \bibinfo {pages}
  {2644} (\bibinfo {year} {2015})}\BibitemShut {NoStop}%
\bibitem [{\citenamefont {Limmer}(2015)}]{limmer_prl115_2015}%
  \BibitemOpen
  \bibfield  {author} {\bibinfo {author} {\bibfnamefont {D.~T.}\ \bibnamefont
  {Limmer}},\ }\href {\doibase 10.1103/PhysRevLett.115.256102} {\bibfield
  {journal} {\bibinfo  {journal} {Phys. Rev. Lett.}\ }\textbf {\bibinfo
  {volume} {115}},\ \bibinfo {pages} {256102} (\bibinfo {year}
  {2015})}\BibitemShut {NoStop}%
\bibitem [{\citenamefont {Uralcan}\ \emph {et~al.}(2016)\citenamefont
  {Uralcan}, \citenamefont {Aksay}, \citenamefont {Debenedetti},\ and\
  \citenamefont {Limmer}}]{uralcan_jpcl7_2016}%
  \BibitemOpen
  \bibfield  {author} {\bibinfo {author} {\bibfnamefont {B.}~\bibnamefont
  {Uralcan}}, \bibinfo {author} {\bibfnamefont {I.~A.}\ \bibnamefont {Aksay}},
  \bibinfo {author} {\bibfnamefont {P.~G.}\ \bibnamefont {Debenedetti}}, \ and\
  \bibinfo {author} {\bibfnamefont {D.~T.}\ \bibnamefont {Limmer}},\ }\href
  {\doibase 10.1021/acs.jpclett.6b00859} {\bibfield  {journal} {\bibinfo
  {journal} {J. Phys. Chem. Lett.}\ }\textbf {\bibinfo {volume} {7}},\ \bibinfo
  {pages} {2333} (\bibinfo {year} {2016})}\BibitemShut {NoStop}%
\bibitem [{\citenamefont {Zhang}\ \emph {et~al.}(2017)\citenamefont {Zhang},
  \citenamefont {Dasbiswas}, \citenamefont {Ludwig}, \citenamefont {Han},
  \citenamefont {Lee}, \citenamefont {Vaikuntanathan},\ and\ \citenamefont
  {Talapin}}]{zhang_nature542_2017}%
  \BibitemOpen
  \bibfield  {author} {\bibinfo {author} {\bibfnamefont {H.}~\bibnamefont
  {Zhang}}, \bibinfo {author} {\bibfnamefont {K.}~\bibnamefont {Dasbiswas}},
  \bibinfo {author} {\bibfnamefont {N.~B.}\ \bibnamefont {Ludwig}}, \bibinfo
  {author} {\bibfnamefont {G.}~\bibnamefont {Han}}, \bibinfo {author}
  {\bibfnamefont {B.}~\bibnamefont {Lee}}, \bibinfo {author} {\bibfnamefont
  {S.}~\bibnamefont {Vaikuntanathan}}, \ and\ \bibinfo {author} {\bibfnamefont
  {D.~V.}\ \bibnamefont {Talapin}},\ }\href {\doibase 10.1038/nature21041}
  {\bibfield  {journal} {\bibinfo  {journal} {Nature}\ }\textbf {\bibinfo
  {volume} {542}},\ \bibinfo {pages} {328} (\bibinfo {year}
  {2017})}\BibitemShut {NoStop}%
\bibitem [{\citenamefont {Attard}(2007)}]{attard_inbook_2007}%
  \BibitemOpen
  \bibfield  {author} {\bibinfo {author} {\bibfnamefont {P.}~\bibnamefont
  {Attard}},\ }\enquote {\bibinfo {title} {Electrolytes and the electric double
  layer},}\ in\ \href {\doibase 10.1002/9780470141519.ch1} {\emph {\bibinfo
  {booktitle} {Advances in Chemical Physics}}},\ Vol.~\bibinfo {volume} {92}\
  (\bibinfo  {publisher} {Wiley-Blackwell},\ \bibinfo {year} {2007})\
  Chap.~\bibinfo {chapter} {1}, pp.\ \bibinfo {pages} {1--159}\BibitemShut
  {NoStop}%
\bibitem [{\citenamefont {Levin}(2002)}]{levin_rpp65_2002}%
  \BibitemOpen
  \bibfield  {author} {\bibinfo {author} {\bibfnamefont {Y.}~\bibnamefont
  {Levin}},\ }\href {\doibase 10.1088/0034-4885/65/11/201} {\bibfield
  {journal} {\bibinfo  {journal} {Rep. Prog. Phys.}\ }\textbf {\bibinfo
  {volume} {65}},\ \bibinfo {pages} {1577} (\bibinfo {year}
  {2002})}\BibitemShut {NoStop}%
\bibitem [{\citenamefont {Evans}\ \emph {et~al.}(1994)\citenamefont {Evans},
  \citenamefont {Leote~de Carvalho}, \citenamefont {Henderson},\ and\
  \citenamefont {Hoyle}}]{evans_jcp100_1994}%
  \BibitemOpen
  \bibfield  {author} {\bibinfo {author} {\bibfnamefont {R.}~\bibnamefont
  {Evans}}, \bibinfo {author} {\bibfnamefont {R.~J.~F.}\ \bibnamefont {Leote~de
  Carvalho}}, \bibinfo {author} {\bibfnamefont {J.~R.}\ \bibnamefont
  {Henderson}}, \ and\ \bibinfo {author} {\bibfnamefont {D.~C.}\ \bibnamefont
  {Hoyle}},\ }\href {\doibase 10.1063/1.466920} {\bibfield  {journal} {\bibinfo
   {journal} {J. Chem. Phys.}\ }\textbf {\bibinfo {volume} {100}},\ \bibinfo
  {pages} {591} (\bibinfo {year} {1994})}\BibitemShut {NoStop}%
\bibitem [{\citenamefont {Attard}(1993)}]{attard_pre48_1993}%
  \BibitemOpen
  \bibfield  {author} {\bibinfo {author} {\bibfnamefont {P.}~\bibnamefont
  {Attard}},\ }\href {\doibase 10.1103/PhysRevE.48.3604} {\bibfield  {journal}
  {\bibinfo  {journal} {Phys. Rev. E}\ }\textbf {\bibinfo {volume} {48}},\
  \bibinfo {pages} {3604} (\bibinfo {year} {1993})}\BibitemShut {NoStop}%
\bibitem [{\citenamefont {Leote~de Carvalho}\ and\ \citenamefont
  {Evans}(1994)}]{leotedecarvalho_mp83_1994}%
  \BibitemOpen
  \bibfield  {author} {\bibinfo {author} {\bibfnamefont {R.~J.~F.}\
  \bibnamefont {Leote~de Carvalho}}\ and\ \bibinfo {author} {\bibfnamefont
  {R.}~\bibnamefont {Evans}},\ }\href {\doibase 10.1080/00268979400101491}
  {\bibfield  {journal} {\bibinfo  {journal} {Mol. Phys.}\ }\textbf {\bibinfo
  {volume} {83}},\ \bibinfo {pages} {619} (\bibinfo {year} {1994})}\BibitemShut
  {NoStop}%
\bibitem [{\citenamefont {Zeng}\ and\ \citenamefont {von
  Klitzing}(2012)}]{zeng_langmuir28_2012}%
  \BibitemOpen
  \bibfield  {author} {\bibinfo {author} {\bibfnamefont {Y.}~\bibnamefont
  {Zeng}}\ and\ \bibinfo {author} {\bibfnamefont {R.}~\bibnamefont {von
  Klitzing}},\ }\href {\doibase 10.1021/la2049822} {\bibfield  {journal}
  {\bibinfo  {journal} {Langmuir}\ }\textbf {\bibinfo {volume} {28}},\ \bibinfo
  {pages} {6313} (\bibinfo {year} {2012})}\BibitemShut {NoStop}%
\bibitem [{\citenamefont {Gebbie}\ \emph {et~al.}(2015)\citenamefont {Gebbie},
  \citenamefont {Dobbs}, \citenamefont {Valtiner},\ and\ \citenamefont
  {Israelachvili}}]{gebbie_pnas112_2015}%
  \BibitemOpen
  \bibfield  {author} {\bibinfo {author} {\bibfnamefont {M.~A.}\ \bibnamefont
  {Gebbie}}, \bibinfo {author} {\bibfnamefont {H.~A.}\ \bibnamefont {Dobbs}},
  \bibinfo {author} {\bibfnamefont {M.}~\bibnamefont {Valtiner}}, \ and\
  \bibinfo {author} {\bibfnamefont {J.~N.}\ \bibnamefont {Israelachvili}},\
  }\href {\doibase 10.1073/pnas.1508366112} {\bibfield  {journal} {\bibinfo
  {journal} {Proc. Natl. Acad. Sci. USA}\ }\textbf {\bibinfo {volume} {112}},\
  \bibinfo {pages} {7432} (\bibinfo {year} {2015})}\BibitemShut {NoStop}%
\bibitem [{\citenamefont {Smith}\ \emph {et~al.}(2017)\citenamefont {Smith},
  \citenamefont {Lee},\ and\ \citenamefont {Perkin}}]{smith_prl118_2017}%
  \BibitemOpen
  \bibfield  {author} {\bibinfo {author} {\bibfnamefont {A.~M.}\ \bibnamefont
  {Smith}}, \bibinfo {author} {\bibfnamefont {A.~A.}\ \bibnamefont {Lee}}, \
  and\ \bibinfo {author} {\bibfnamefont {S.}~\bibnamefont {Perkin}},\ }\href
  {\doibase 10.1103/PhysRevLett.118.096002} {\bibfield  {journal} {\bibinfo
  {journal} {Phys. Rev. Lett.}\ }\textbf {\bibinfo {volume} {118}},\ \bibinfo
  {pages} {096002} (\bibinfo {year} {2017})}\BibitemShut {NoStop}%
\bibitem [{\citenamefont {McEwen}\ \emph {et~al.}(1999)\citenamefont {McEwen},
  \citenamefont {Ngo}, \citenamefont {LeCompte},\ and\ \citenamefont
  {Goldman}}]{mcewen_jes146_1999}%
  \BibitemOpen
  \bibfield  {author} {\bibinfo {author} {\bibfnamefont {A.~B.}\ \bibnamefont
  {McEwen}}, \bibinfo {author} {\bibfnamefont {H.~L.}\ \bibnamefont {Ngo}},
  \bibinfo {author} {\bibfnamefont {K.}~\bibnamefont {LeCompte}}, \ and\
  \bibinfo {author} {\bibfnamefont {J.~L.}\ \bibnamefont {Goldman}},\ }\href
  {\doibase 10.1149/1.1391827} {\bibfield  {journal} {\bibinfo  {journal} {J.
  Electrochem. Soc.}\ }\textbf {\bibinfo {volume} {146}},\ \bibinfo {pages}
  {1687} (\bibinfo {year} {1999})}\BibitemShut {NoStop}%
\bibitem [{\citenamefont {Zhu}\ \emph {et~al.}(2011)\citenamefont {Zhu},
  \citenamefont {Murali}, \citenamefont {Stoller}, \citenamefont {Ganesh},
  \citenamefont {Cai}, \citenamefont {Ferreira}, \citenamefont {Pirkle},
  \citenamefont {Wallace}, \citenamefont {Cychosz}, \citenamefont {Thommes},
  \citenamefont {Su}, \citenamefont {Stach},\ and\ \citenamefont
  {Ruoff}}]{zhu_science332_2011}%
  \BibitemOpen
  \bibfield  {author} {\bibinfo {author} {\bibfnamefont {Y.}~\bibnamefont
  {Zhu}}, \bibinfo {author} {\bibfnamefont {S.}~\bibnamefont {Murali}},
  \bibinfo {author} {\bibfnamefont {M.~D.}\ \bibnamefont {Stoller}}, \bibinfo
  {author} {\bibfnamefont {K.~J.}\ \bibnamefont {Ganesh}}, \bibinfo {author}
  {\bibfnamefont {W.}~\bibnamefont {Cai}}, \bibinfo {author} {\bibfnamefont
  {P.~J.}\ \bibnamefont {Ferreira}}, \bibinfo {author} {\bibfnamefont
  {A.}~\bibnamefont {Pirkle}}, \bibinfo {author} {\bibfnamefont {R.~M.}\
  \bibnamefont {Wallace}}, \bibinfo {author} {\bibfnamefont {K.~A.}\
  \bibnamefont {Cychosz}}, \bibinfo {author} {\bibfnamefont {M.}~\bibnamefont
  {Thommes}}, \bibinfo {author} {\bibfnamefont {D.}~\bibnamefont {Su}},
  \bibinfo {author} {\bibfnamefont {E.~A.}\ \bibnamefont {Stach}}, \ and\
  \bibinfo {author} {\bibfnamefont {R.~S.}\ \bibnamefont {Ruoff}},\ }\href
  {\doibase 10.1126/science.1200770} {\bibfield  {journal} {\bibinfo  {journal}
  {Science}\ }\textbf {\bibinfo {volume} {332}},\ \bibinfo {pages} {1537}
  (\bibinfo {year} {2011})}\BibitemShut {NoStop}%
\bibitem [{\citenamefont {Yang}\ \emph {et~al.}(2013)\citenamefont {Yang},
  \citenamefont {Cheng}, \citenamefont {Wang}, \citenamefont {Qiu},\ and\
  \citenamefont {Li}}]{yang_science341_2013}%
  \BibitemOpen
  \bibfield  {author} {\bibinfo {author} {\bibfnamefont {X.}~\bibnamefont
  {Yang}}, \bibinfo {author} {\bibfnamefont {C.}~\bibnamefont {Cheng}},
  \bibinfo {author} {\bibfnamefont {Y.}~\bibnamefont {Wang}}, \bibinfo {author}
  {\bibfnamefont {L.}~\bibnamefont {Qiu}}, \ and\ \bibinfo {author}
  {\bibfnamefont {D.}~\bibnamefont {Li}},\ }\href {\doibase
  10.1126/science.1239089} {\bibfield  {journal} {\bibinfo  {journal}
  {Science}\ }\textbf {\bibinfo {volume} {341}},\ \bibinfo {pages} {534}
  (\bibinfo {year} {2013})}\BibitemShut {NoStop}%
\bibitem [{\citenamefont {Moazzami-Gudarzi}\ \emph {et~al.}(2016)\citenamefont
  {Moazzami-Gudarzi}, \citenamefont {Kremer}, \citenamefont {Valmacco},
  \citenamefont {Maroni}, \citenamefont {Borkovec},\ and\ \citenamefont
  {Trefalt}}]{moazzami-gudarzi_prl117_2016}%
  \BibitemOpen
  \bibfield  {author} {\bibinfo {author} {\bibfnamefont {M.}~\bibnamefont
  {Moazzami-Gudarzi}}, \bibinfo {author} {\bibfnamefont {T.}~\bibnamefont
  {Kremer}}, \bibinfo {author} {\bibfnamefont {V.}~\bibnamefont {Valmacco}},
  \bibinfo {author} {\bibfnamefont {P.}~\bibnamefont {Maroni}}, \bibinfo
  {author} {\bibfnamefont {M.}~\bibnamefont {Borkovec}}, \ and\ \bibinfo
  {author} {\bibfnamefont {G.}~\bibnamefont {Trefalt}},\ }\href {\doibase
  10.1103/PhysRevLett.117.088001} {\bibfield  {journal} {\bibinfo  {journal}
  {Phys. Rev. Lett.}\ }\textbf {\bibinfo {volume} {117}},\ \bibinfo {pages}
  {088001} (\bibinfo {year} {2016})}\BibitemShut {NoStop}%
\bibitem [{\citenamefont {Sch\"on}\ and\ \citenamefont {von
  Klitzing}(2018)}]{schoen_bjnt9_2018}%
  \BibitemOpen
  \bibfield  {author} {\bibinfo {author} {\bibfnamefont {S.}~\bibnamefont
  {Sch\"on}}\ and\ \bibinfo {author} {\bibfnamefont {R.}~\bibnamefont {von
  Klitzing}},\ }\href {\doibase 10.3762/bjnano.9.101} {\bibfield  {journal}
  {\bibinfo  {journal} {Beilstein J. Nanotechnol.}\ }\textbf {\bibinfo {volume}
  {9}},\ \bibinfo {pages} {1095} (\bibinfo {year} {2018})}\BibitemShut
  {NoStop}%
\bibitem [{\citenamefont {Grodon}\ \emph {et~al.}(2004)\citenamefont {Grodon},
  \citenamefont {Dijkstra}, \citenamefont {Evans},\ and\ \citenamefont
  {Roth}}]{grodon_jcp121_2004}%
  \BibitemOpen
  \bibfield  {author} {\bibinfo {author} {\bibfnamefont {C.}~\bibnamefont
  {Grodon}}, \bibinfo {author} {\bibfnamefont {M.}~\bibnamefont {Dijkstra}},
  \bibinfo {author} {\bibfnamefont {R.}~\bibnamefont {Evans}}, \ and\ \bibinfo
  {author} {\bibfnamefont {R.}~\bibnamefont {Roth}},\ }\href {\doibase
  10.1063/1.1798057} {\bibfield  {journal} {\bibinfo  {journal} {J. Chem.
  Phys.}\ }\textbf {\bibinfo {volume} {121}},\ \bibinfo {pages} {7869}
  (\bibinfo {year} {2004})}\BibitemShut {NoStop}%
\bibitem [{\citenamefont {Baumgartl}\ \emph {et~al.}(2007)\citenamefont
  {Baumgartl}, \citenamefont {Dullens}, \citenamefont {Dijkstra}, \citenamefont
  {Roth},\ and\ \citenamefont {Bechinger}}]{baumgartl_prl98_2007}%
  \BibitemOpen
  \bibfield  {author} {\bibinfo {author} {\bibfnamefont {J.}~\bibnamefont
  {Baumgartl}}, \bibinfo {author} {\bibfnamefont {R.~P.~A.}\ \bibnamefont
  {Dullens}}, \bibinfo {author} {\bibfnamefont {M.}~\bibnamefont {Dijkstra}},
  \bibinfo {author} {\bibfnamefont {R.}~\bibnamefont {Roth}}, \ and\ \bibinfo
  {author} {\bibfnamefont {C.}~\bibnamefont {Bechinger}},\ }\href {\doibase
  10.1103/PhysRevLett.98.198303} {\bibfield  {journal} {\bibinfo  {journal}
  {Phys. Rev. Lett.}\ }\textbf {\bibinfo {volume} {98}},\ \bibinfo {pages}
  {198303} (\bibinfo {year} {2007})}\BibitemShut {NoStop}%
\bibitem [{\citenamefont {Statt}\ \emph {et~al.}(2016)\citenamefont {Statt},
  \citenamefont {Pinchaipat}, \citenamefont {Turci}, \citenamefont {Evans},\
  and\ \citenamefont {Royall}}]{statt_jcp144_2016}%
  \BibitemOpen
  \bibfield  {author} {\bibinfo {author} {\bibfnamefont {A.}~\bibnamefont
  {Statt}}, \bibinfo {author} {\bibfnamefont {R.}~\bibnamefont {Pinchaipat}},
  \bibinfo {author} {\bibfnamefont {F.}~\bibnamefont {Turci}}, \bibinfo
  {author} {\bibfnamefont {R.}~\bibnamefont {Evans}}, \ and\ \bibinfo {author}
  {\bibfnamefont {C.~P.}\ \bibnamefont {Royall}},\ }\href {\doibase
  10.1063/1.4945808} {\bibfield  {journal} {\bibinfo  {journal} {J. Chem.
  Phys.}\ }\textbf {\bibinfo {volume} {144}},\ \bibinfo {pages} {144506}
  (\bibinfo {year} {2016})}\BibitemShut {NoStop}%
\bibitem [{\citenamefont {Grimson}\ and\ \citenamefont
  {Rickayzen}(1982)}]{grimson_cpl86_1982}%
  \BibitemOpen
  \bibfield  {author} {\bibinfo {author} {\bibfnamefont {M.~J.}\ \bibnamefont
  {Grimson}}\ and\ \bibinfo {author} {\bibfnamefont {G.}~\bibnamefont
  {Rickayzen}},\ }\href {\doibase 10.1016/0009-2614(82)83119-9} {\bibfield
  {journal} {\bibinfo  {journal} {Chem. Phys. Lett.}\ }\textbf {\bibinfo
  {volume} {86}},\ \bibinfo {pages} {71} (\bibinfo {year} {1982})}\BibitemShut
  {NoStop}%
\bibitem [{\citenamefont {Tang}\ \emph {et~al.}(1992)\citenamefont {Tang},
  \citenamefont {Scriven},\ and\ \citenamefont {Davis}}]{tang_jcp97_1992}%
  \BibitemOpen
  \bibfield  {author} {\bibinfo {author} {\bibfnamefont {Z.}~\bibnamefont
  {Tang}}, \bibinfo {author} {\bibfnamefont {L.~E.}\ \bibnamefont {Scriven}}, \
  and\ \bibinfo {author} {\bibfnamefont {H.~T.}\ \bibnamefont {Davis}},\ }\href
  {\doibase 10.1063/1.463595} {\bibfield  {journal} {\bibinfo  {journal} {J.
  Chem. Phys.}\ }\textbf {\bibinfo {volume} {97}},\ \bibinfo {pages} {494}
  (\bibinfo {year} {1992})}\BibitemShut {NoStop}%
\bibitem [{\citenamefont {Boda}\ and\ \citenamefont
  {Henderson}(2000)}]{boda_jcp112_2000}%
  \BibitemOpen
  \bibfield  {author} {\bibinfo {author} {\bibfnamefont {D.}~\bibnamefont
  {Boda}}\ and\ \bibinfo {author} {\bibfnamefont {D.}~\bibnamefont
  {Henderson}},\ }\href {\doibase 10.1063/1.481507} {\bibfield  {journal}
  {\bibinfo  {journal} {J. Chem. Phys.}\ }\textbf {\bibinfo {volume} {112}},\
  \bibinfo {pages} {8934} (\bibinfo {year} {2000})}\BibitemShut {NoStop}%
\bibitem [{\citenamefont {Rotenberg}\ \emph {et~al.}(2018)\citenamefont
  {Rotenberg}, \citenamefont {Bernard},\ and\ \citenamefont
  {Hansen}}]{rotenberg_jpcm30_2018}%
  \BibitemOpen
  \bibfield  {author} {\bibinfo {author} {\bibfnamefont {B.}~\bibnamefont
  {Rotenberg}}, \bibinfo {author} {\bibfnamefont {O.}~\bibnamefont {Bernard}},
  \ and\ \bibinfo {author} {\bibfnamefont {J.-P.}\ \bibnamefont {Hansen}},\
  }\href {\doibase 10.1088/1361-648X/aaa3ac} {\bibfield  {journal} {\bibinfo
  {journal} {J. Phys.: Condens. Matter}\ }\textbf {\bibinfo {volume} {30}},\
  \bibinfo {pages} {054005} (\bibinfo {year} {2018})}\BibitemShut {NoStop}%
\bibitem [{\citenamefont {Limbach}\ \emph {et~al.}(2006)\citenamefont
  {Limbach}, \citenamefont {Arnold}, \citenamefont {Mann},\ and\ \citenamefont
  {Holm}}]{limbach_cpc174_2006}%
  \BibitemOpen
  \bibfield  {author} {\bibinfo {author} {\bibfnamefont {H.~J.}\ \bibnamefont
  {Limbach}}, \bibinfo {author} {\bibfnamefont {A.}~\bibnamefont {Arnold}},
  \bibinfo {author} {\bibfnamefont {B.~A.}\ \bibnamefont {Mann}}, \ and\
  \bibinfo {author} {\bibfnamefont {C.}~\bibnamefont {Holm}},\ }\href {\doibase
  10.1016/j.cpc.2005.10.005} {\bibfield  {journal} {\bibinfo  {journal} {Comp.
  Phys. Comm.}\ }\textbf {\bibinfo {volume} {174}},\ \bibinfo {pages} {704}
  (\bibinfo {year} {2006})}\BibitemShut {NoStop}%
\bibitem [{\citenamefont {Arnold}\ \emph {et~al.}(2013)\citenamefont {Arnold},
  \citenamefont {Lenz}, \citenamefont {Kesselheim}, \citenamefont {Weeber},
  \citenamefont {Fahrenberger}, \citenamefont {Roehm}, \citenamefont
  {Ko\v{s}ovan},\ and\ \citenamefont {Holm}}]{arnold_book_2013}%
  \BibitemOpen
  \bibfield  {author} {\bibinfo {author} {\bibfnamefont {A.}~\bibnamefont
  {Arnold}}, \bibinfo {author} {\bibfnamefont {O.}~\bibnamefont {Lenz}},
  \bibinfo {author} {\bibfnamefont {S.}~\bibnamefont {Kesselheim}}, \bibinfo
  {author} {\bibfnamefont {R.}~\bibnamefont {Weeber}}, \bibinfo {author}
  {\bibfnamefont {F.}~\bibnamefont {Fahrenberger}}, \bibinfo {author}
  {\bibfnamefont {D.}~\bibnamefont {Roehm}}, \bibinfo {author} {\bibfnamefont
  {P.}~\bibnamefont {Ko\v{s}ovan}}, \ and\ \bibinfo {author} {\bibfnamefont
  {C.}~\bibnamefont {Holm}},\ }in\ \href {\doibase 10.1007/978-3-642-32979-1_1}
  {\emph {\bibinfo {booktitle} {Meshfree Methods for Partial Differential
  Equations {VI}}}},\ \bibinfo {series} {Lecture Notes in Computational Science
  and Engineering}, Vol.~\bibinfo {volume} {89},\ \bibinfo {editor} {edited by\
  \bibinfo {editor} {\bibfnamefont {M.}~\bibnamefont {Griebel}}\ and\ \bibinfo
  {editor} {\bibfnamefont {M.~A.}\ \bibnamefont {Schweitzer}}}\ (\bibinfo
  {publisher} {Springer},\ \bibinfo {year} {2013})\ pp.\ \bibinfo {pages}
  {1--23}\BibitemShut {NoStop}%
\bibitem [{\citenamefont {Hansen}\ and\ \citenamefont
  {McDonald}(2013)}]{hansen_book_2013}%
  \BibitemOpen
  \bibfield  {author} {\bibinfo {author} {\bibfnamefont {J.-P.}\ \bibnamefont
  {Hansen}}\ and\ \bibinfo {author} {\bibfnamefont {I.~R.}\ \bibnamefont
  {McDonald}},\ }\href@noop {} {\emph {\bibinfo {title} {Theory of simple
  liquids}}},\ \bibinfo {edition} {4th}\ ed.\ (\bibinfo  {publisher}
  {Elsevier},\ \bibinfo {address} {~},\ \bibinfo {year} {2013})\BibitemShut
  {NoStop}%
\bibitem [{\citenamefont {Kjellander}\ and\ \citenamefont
  {Mitchell}(1992)}]{kjellander_cpl200_1992}%
  \BibitemOpen
  \bibfield  {author} {\bibinfo {author} {\bibfnamefont {R.}~\bibnamefont
  {Kjellander}}\ and\ \bibinfo {author} {\bibfnamefont {D.}~\bibnamefont
  {Mitchell}},\ }\href {\doibase 10.1016/0009-2614(92)87048-T} {\bibfield
  {journal} {\bibinfo  {journal} {Chem. Phys. Lett.}\ }\textbf {\bibinfo
  {volume} {200}},\ \bibinfo {pages} {76} (\bibinfo {year} {1992})}\BibitemShut
  {NoStop}%
\bibitem [{\citenamefont {Fisher}\ and\ \citenamefont
  {Wiodm}(1969)}]{fisher_jcp50_1969}%
  \BibitemOpen
  \bibfield  {author} {\bibinfo {author} {\bibfnamefont {M.~E.}\ \bibnamefont
  {Fisher}}\ and\ \bibinfo {author} {\bibfnamefont {B.}~\bibnamefont {Wiodm}},\
  }\href {\doibase 10.1063/1.1671624} {\bibfield  {journal} {\bibinfo
  {journal} {J. Chem. Phys.}\ }\textbf {\bibinfo {volume} {50}},\ \bibinfo
  {pages} {3756} (\bibinfo {year} {1969})}\BibitemShut {NoStop}%
\bibitem [{\citenamefont {Henderson}\ and\ \citenamefont
  {Sabeur}(1992)}]{henderson_jcp97_1992}%
  \BibitemOpen
  \bibfield  {author} {\bibinfo {author} {\bibfnamefont {J.~R.}\ \bibnamefont
  {Henderson}}\ and\ \bibinfo {author} {\bibfnamefont {Z.~A.}\ \bibnamefont
  {Sabeur}},\ }\href {\doibase 10.1063/1.463652} {\bibfield  {journal}
  {\bibinfo  {journal} {J. Chem. Phys.}\ }\textbf {\bibinfo {volume} {97}},\
  \bibinfo {pages} {6750} (\bibinfo {year} {1992})}\BibitemShut {NoStop}%
\bibitem [{\citenamefont {Hansen-Goos}\ and\ \citenamefont
  {Roth}(2006)}]{hansen-goos_jpcm18_2006}%
  \BibitemOpen
  \bibfield  {author} {\bibinfo {author} {\bibfnamefont {H.}~\bibnamefont
  {Hansen-Goos}}\ and\ \bibinfo {author} {\bibfnamefont {R.}~\bibnamefont
  {Roth}},\ }\href {\doibase 10.1088/0953-8984/18/37/002} {\bibfield  {journal}
  {\bibinfo  {journal} {J. Phys.: Condens. Matter}\ }\textbf {\bibinfo {volume}
  {18}},\ \bibinfo {pages} {8413} (\bibinfo {year} {2006})}\BibitemShut
  {NoStop}%
\bibitem [{\citenamefont {Kirkwood}(1939)}]{kirkwood_jcp7_1939}%
  \BibitemOpen
  \bibfield  {author} {\bibinfo {author} {\bibfnamefont {J.~G.}\ \bibnamefont
  {Kirkwood}},\ }\href {\doibase 10.1063/1.1750344} {\bibfield  {journal}
  {\bibinfo  {journal} {J. Chem. Phys.}\ }\textbf {\bibinfo {volume} {7}},\
  \bibinfo {pages} {919} (\bibinfo {year} {1939})}\BibitemShut {NoStop}%
\bibitem [{SM()}]{SM}%
  \BibitemOpen
  \href@noop {} {}\bibinfo {note} {See Supplemental Material at
  \url{http://link.aps.org/supplemental/10.1103/PhysRevLett.121.075501} for a
  PDF document containing simulation results for asymmetric ions and a
  comparison between theory and experimental data. The document includes
  Refs.~\cite{smith_prl118_2017,smith_jpcl7_2016,coles_fd206_2018,santos_jcp134_2011,soetens_jml92_2001,oettel_pre82_2010}}\BibitemShut
  {NoStop}%
\bibitem [{\citenamefont {Parrinello}\ and\ \citenamefont
  {Tosi}(1979)}]{parrinello_rnc2_1979}%
  \BibitemOpen
  \bibfield  {author} {\bibinfo {author} {\bibfnamefont {M.}~\bibnamefont
  {Parrinello}}\ and\ \bibinfo {author} {\bibfnamefont {M.~P.}\ \bibnamefont
  {Tosi}},\ }\href {\doibase 10.1007/BF02724355} {\bibfield  {journal}
  {\bibinfo  {journal} {Riv. Nuovo Cim.}\ }\textbf {\bibinfo {volume} {2}},\
  \bibinfo {pages} {1} (\bibinfo {year} {1979})}\BibitemShut {NoStop}%
\bibitem [{\citenamefont {Stell}\ \emph {et~al.}(1976)\citenamefont {Stell},
  \citenamefont {Wu},\ and\ \citenamefont {Larsen}}]{stell_prl37_1976}%
  \BibitemOpen
  \bibfield  {author} {\bibinfo {author} {\bibfnamefont {G.}~\bibnamefont
  {Stell}}, \bibinfo {author} {\bibfnamefont {K.~C.}\ \bibnamefont {Wu}}, \
  and\ \bibinfo {author} {\bibfnamefont {B.}~\bibnamefont {Larsen}},\ }\href
  {\doibase 10.1103/PhysRevLett.37.1369} {\bibfield  {journal} {\bibinfo
  {journal} {Phys. Rev. Lett.}\ }\textbf {\bibinfo {volume} {37}},\ \bibinfo
  {pages} {1369} (\bibinfo {year} {1976})}\BibitemShut {NoStop}%
\bibitem [{\citenamefont {Lee}\ \emph {et~al.}(2017)\citenamefont {Lee},
  \citenamefont {Perez-Martinez}, \citenamefont {Smith},\ and\ \citenamefont
  {Perkin}}]{lee_prl119_2017}%
  \BibitemOpen
  \bibfield  {author} {\bibinfo {author} {\bibfnamefont {A.~A.}\ \bibnamefont
  {Lee}}, \bibinfo {author} {\bibfnamefont {C.~S.}\ \bibnamefont
  {Perez-Martinez}}, \bibinfo {author} {\bibfnamefont {A.~M.}\ \bibnamefont
  {Smith}}, \ and\ \bibinfo {author} {\bibfnamefont {S.}~\bibnamefont
  {Perkin}},\ }\href {\doibase 10.1103/PhysRevLett.119.026002} {\bibfield
  {journal} {\bibinfo  {journal} {Phys. Rev. Lett.}\ }\textbf {\bibinfo
  {volume} {119}},\ \bibinfo {pages} {026002} (\bibinfo {year}
  {2017})}\BibitemShut {NoStop}%
\bibitem [{\citenamefont {Evans}\ \emph {et~al.}(1993)\citenamefont {Evans},
  \citenamefont {Henderson}, \citenamefont {Hoyle}, \citenamefont {Parry},\
  and\ \citenamefont {Sabeur}}]{evans_mp80_1993}%
  \BibitemOpen
  \bibfield  {author} {\bibinfo {author} {\bibfnamefont {R.}~\bibnamefont
  {Evans}}, \bibinfo {author} {\bibfnamefont {J.~R.}\ \bibnamefont
  {Henderson}}, \bibinfo {author} {\bibfnamefont {D.~C.}\ \bibnamefont
  {Hoyle}}, \bibinfo {author} {\bibfnamefont {A.~O.}\ \bibnamefont {Parry}}, \
  and\ \bibinfo {author} {\bibfnamefont {Z.~A.}\ \bibnamefont {Sabeur}},\
  }\href {\doibase 10.1080/00268979300102621} {\bibfield  {journal} {\bibinfo
  {journal} {Mol. Phys.}\ }\textbf {\bibinfo {volume} {80}},\ \bibinfo {pages}
  {755} (\bibinfo {year} {1993})}\BibitemShut {NoStop}%
\bibitem [{\citenamefont {Gottwald}\ \emph {et~al.}(2004)\citenamefont
  {Gottwald}, \citenamefont {Likos}, \citenamefont {Kahl},\ and\ \citenamefont
  {L\"owen}}]{gottwald_prl92_2004}%
  \BibitemOpen
  \bibfield  {author} {\bibinfo {author} {\bibfnamefont {D.}~\bibnamefont
  {Gottwald}}, \bibinfo {author} {\bibfnamefont {C.~N.}\ \bibnamefont {Likos}},
  \bibinfo {author} {\bibfnamefont {G.}~\bibnamefont {Kahl}}, \ and\ \bibinfo
  {author} {\bibfnamefont {H.}~\bibnamefont {L\"owen}},\ }\href {\doibase
  10.1103/PhysRevLett.92.068301} {\bibfield  {journal} {\bibinfo  {journal}
  {Phys. Rev. Lett.}\ }\textbf {\bibinfo {volume} {92}},\ \bibinfo {pages}
  {068301} (\bibinfo {year} {2004})}\BibitemShut {NoStop}%
\bibitem [{\citenamefont {L\'eger}\ and\ \citenamefont
  {Levesque}(2005)}]{leger_jcp123_2005}%
  \BibitemOpen
  \bibfield  {author} {\bibinfo {author} {\bibfnamefont {D.}~\bibnamefont
  {L\'eger}}\ and\ \bibinfo {author} {\bibfnamefont {D.}~\bibnamefont
  {Levesque}},\ }\href {\doibase 10.1063/1.1979480} {\bibfield  {journal}
  {\bibinfo  {journal} {J. Chem. Phys.}\ }\textbf {\bibinfo {volume} {123}},\
  \bibinfo {pages} {124910} (\bibinfo {year} {2005})}\BibitemShut {NoStop}%
\bibitem [{\citenamefont {Denton}(2017)}]{denton_pre96_2017}%
  \BibitemOpen
  \bibfield  {author} {\bibinfo {author} {\bibfnamefont {A.~R.}\ \bibnamefont
  {Denton}},\ }\href {\doibase 10.1103/PhysRevE.96.062610} {\bibfield
  {journal} {\bibinfo  {journal} {Phys. Rev. E}\ }\textbf {\bibinfo {volume}
  {96}},\ \bibinfo {pages} {062610} (\bibinfo {year} {2017})}\BibitemShut
  {NoStop}%
\bibitem [{\citenamefont {Chernov}\ and\ \citenamefont
  {Mikheev}(1988)}]{chernov_prl60_1988}%
  \BibitemOpen
  \bibfield  {author} {\bibinfo {author} {\bibfnamefont {A.~A.}\ \bibnamefont
  {Chernov}}\ and\ \bibinfo {author} {\bibfnamefont {L.~V.}\ \bibnamefont
  {Mikheev}},\ }\href {\doibase 10.1103/PhysRevLett.60.2488} {\bibfield
  {journal} {\bibinfo  {journal} {Phys. Rev. Lett.}\ }\textbf {\bibinfo
  {volume} {60}},\ \bibinfo {pages} {2488} (\bibinfo {year}
  {1988})}\BibitemShut {NoStop}%
\bibitem [{\citenamefont {Henderson}(1994)}]{henderson_pre50_1994}%
  \BibitemOpen
  \bibfield  {author} {\bibinfo {author} {\bibfnamefont {J.~R.}\ \bibnamefont
  {Henderson}},\ }\href {\doibase 10.1103/PhysRevE.50.4836} {\bibfield
  {journal} {\bibinfo  {journal} {Phys. Rev. E}\ }\textbf {\bibinfo {volume}
  {50}},\ \bibinfo {pages} {4836} (\bibinfo {year} {1994})}\BibitemShut
  {NoStop}%
\bibitem [{\citenamefont {Li}\ \emph {et~al.}(2017)\citenamefont {Li},
  \citenamefont {Girard}, \citenamefont {Shen}, \citenamefont {Millan},\ and\
  \citenamefont {Olvera de~la Cruz}}]{li_pnas114_2017}%
  \BibitemOpen
  \bibfield  {author} {\bibinfo {author} {\bibfnamefont {Y.}~\bibnamefont
  {Li}}, \bibinfo {author} {\bibfnamefont {M.}~\bibnamefont {Girard}}, \bibinfo
  {author} {\bibfnamefont {M.}~\bibnamefont {Shen}}, \bibinfo {author}
  {\bibfnamefont {J.~A.}\ \bibnamefont {Millan}}, \ and\ \bibinfo {author}
  {\bibfnamefont {M.}~\bibnamefont {Olvera de~la Cruz}},\ }\href {\doibase
  10.1073/pnas.1713168114} {\bibfield  {journal} {\bibinfo  {journal} {Proc.
  Natl. Acad. Sci. USA}\ }\textbf {\bibinfo {volume} {114}},\ \bibinfo {pages}
  {11838} (\bibinfo {year} {2017})}\BibitemShut {NoStop}%
\bibitem [{\citenamefont {Morfill}\ and\ \citenamefont
  {Ivlev}(2009)}]{morfill_rmp81_2009}%
  \BibitemOpen
  \bibfield  {author} {\bibinfo {author} {\bibfnamefont {G.~E.}\ \bibnamefont
  {Morfill}}\ and\ \bibinfo {author} {\bibfnamefont {A.~V.}\ \bibnamefont
  {Ivlev}},\ }\href {\doibase 10.1103/RevModPhys.81.1353} {\bibfield  {journal}
  {\bibinfo  {journal} {Rev. Mod. Phys.}\ }\textbf {\bibinfo {volume} {81}},\
  \bibinfo {pages} {1353} (\bibinfo {year} {2009})}\BibitemShut {NoStop}%
\bibitem [{\citenamefont {Brader}\ \emph {et~al.}(2001)\citenamefont {Brader},
  \citenamefont {Dijkstra},\ and\ \citenamefont {Evans}}]{brader_pre63_2001}%
  \BibitemOpen
  \bibfield  {author} {\bibinfo {author} {\bibfnamefont {J.~M.}\ \bibnamefont
  {Brader}}, \bibinfo {author} {\bibfnamefont {M.}~\bibnamefont {Dijkstra}}, \
  and\ \bibinfo {author} {\bibfnamefont {R.}~\bibnamefont {Evans}},\ }\href
  {\doibase 10.1103/PhysRevE.63.041405} {\bibfield  {journal} {\bibinfo
  {journal} {Phys. Rev. E}\ }\textbf {\bibinfo {volume} {63}},\ \bibinfo
  {pages} {041405} (\bibinfo {year} {2001})}\BibitemShut {NoStop}%
\bibitem [{\citenamefont {Archer}\ \emph {et~al.}(2007)\citenamefont {Archer},
  \citenamefont {Pini}, \citenamefont {Evans},\ and\ \citenamefont
  {Reatto}}]{archer_jcp126_2007}%
  \BibitemOpen
  \bibfield  {author} {\bibinfo {author} {\bibfnamefont {A.~J.}\ \bibnamefont
  {Archer}}, \bibinfo {author} {\bibfnamefont {D.}~\bibnamefont {Pini}},
  \bibinfo {author} {\bibfnamefont {R.}~\bibnamefont {Evans}}, \ and\ \bibinfo
  {author} {\bibfnamefont {L.}~\bibnamefont {Reatto}},\ }\href {\doibase
  10.1063/1.2405355} {\bibfield  {journal} {\bibinfo  {journal} {J. Chem.
  Phys.}\ }\textbf {\bibinfo {volume} {126}},\ \bibinfo {pages} {014104}
  (\bibinfo {year} {2007})}\BibitemShut {NoStop}%
\bibitem [{\citenamefont {Smith}\ \emph {et~al.}(2016)\citenamefont {Smith},
  \citenamefont {Lee},\ and\ \citenamefont {Perkin}}]{smith_jpcl7_2016}%
  \BibitemOpen
  \bibfield  {author} {\bibinfo {author} {\bibfnamefont {A.~M.}\ \bibnamefont
  {Smith}}, \bibinfo {author} {\bibfnamefont {A.~A.}\ \bibnamefont {Lee}}, \
  and\ \bibinfo {author} {\bibfnamefont {S.}~\bibnamefont {Perkin}},\ }\href
  {\doibase 10.1021/acs.jpclett.6b00867} {\bibfield  {journal} {\bibinfo
  {journal} {J. Phys. Chem. Lett.}\ }\textbf {\bibinfo {volume} {7}},\ \bibinfo
  {pages} {2157} (\bibinfo {year} {2016})}\BibitemShut {NoStop}%
\bibitem [{\citenamefont {Coles}\ \emph {et~al.}(2018)\citenamefont {Coles},
  \citenamefont {Smith}, \citenamefont {Fedorov}, \citenamefont {Hausen},\ and\
  \citenamefont {Perkin}}]{coles_fd206_2018}%
  \BibitemOpen
  \bibfield  {author} {\bibinfo {author} {\bibfnamefont {S.~W.}\ \bibnamefont
  {Coles}}, \bibinfo {author} {\bibfnamefont {A.~M.}\ \bibnamefont {Smith}},
  \bibinfo {author} {\bibfnamefont {M.~V.}\ \bibnamefont {Fedorov}}, \bibinfo
  {author} {\bibfnamefont {F.}~\bibnamefont {Hausen}}, \ and\ \bibinfo {author}
  {\bibfnamefont {S.}~\bibnamefont {Perkin}},\ }\href {\doibase
  10.1039/C7FD00168A} {\bibfield  {journal} {\bibinfo  {journal} {Faraday
  Discuss.}\ }\textbf {\bibinfo {volume} {206}},\ \bibinfo {pages} {427}
  (\bibinfo {year} {2018})}\BibitemShut {NoStop}%
\bibitem [{\citenamefont {Santos}\ \emph {et~al.}(2011)\citenamefont {Santos},
  \citenamefont {Murthy}, \citenamefont {Baker},\ and\ \citenamefont
  {Castner}}]{santos_jcp134_2011}%
  \BibitemOpen
  \bibfield  {author} {\bibinfo {author} {\bibfnamefont {C.~S.}\ \bibnamefont
  {Santos}}, \bibinfo {author} {\bibfnamefont {N.~S.}\ \bibnamefont {Murthy}},
  \bibinfo {author} {\bibfnamefont {G.~A.}\ \bibnamefont {Baker}}, \ and\
  \bibinfo {author} {\bibfnamefont {E.~W.}\ \bibnamefont {Castner}},\ }\href
  {\doibase 10.1063/1.3569131} {\bibfield  {journal} {\bibinfo  {journal} {J.
  Chem. Phys.}\ }\textbf {\bibinfo {volume} {134}},\ \bibinfo {pages} {121101}
  (\bibinfo {year} {2011})}\BibitemShut {NoStop}%
\bibitem [{\citenamefont {Soetens}\ \emph {et~al.}(2001)\citenamefont
  {Soetens}, \citenamefont {Millot}, \citenamefont {Maigret},\ and\
  \citenamefont {Bak\'o}}]{soetens_jml92_2001}%
  \BibitemOpen
  \bibfield  {author} {\bibinfo {author} {\bibfnamefont {J.-C.}\ \bibnamefont
  {Soetens}}, \bibinfo {author} {\bibfnamefont {C.}~\bibnamefont {Millot}},
  \bibinfo {author} {\bibfnamefont {B.}~\bibnamefont {Maigret}}, \ and\
  \bibinfo {author} {\bibfnamefont {I.}~\bibnamefont {Bak\'o}},\ }\href
  {\doibase 10.1016/S0167-7322(01)00192-1} {\bibfield  {journal} {\bibinfo
  {journal} {J. Mol. Liq.}\ }\textbf {\bibinfo {volume} {92}},\ \bibinfo
  {pages} {201} (\bibinfo {year} {2001})}\BibitemShut {NoStop}%
\bibitem [{\citenamefont {Oettel}\ \emph {et~al.}(2010)\citenamefont {Oettel},
  \citenamefont {G\"orig}, \citenamefont {H\"artel}, \citenamefont {L\"owen},
  \citenamefont {Radu},\ and\ \citenamefont {Schilling}}]{oettel_pre82_2010}%
  \BibitemOpen
  \bibfield  {author} {\bibinfo {author} {\bibfnamefont {M.}~\bibnamefont
  {Oettel}}, \bibinfo {author} {\bibfnamefont {S.}~\bibnamefont {G\"orig}},
  \bibinfo {author} {\bibfnamefont {A.}~\bibnamefont {H\"artel}}, \bibinfo
  {author} {\bibfnamefont {H.}~\bibnamefont {L\"owen}}, \bibinfo {author}
  {\bibfnamefont {M.}~\bibnamefont {Radu}}, \ and\ \bibinfo {author}
  {\bibfnamefont {T.}~\bibnamefont {Schilling}},\ }\href {\doibase
  10.1103/PhysRevE.82.051404} {\bibfield  {journal} {\bibinfo  {journal} {Phys.
  Rev. E}\ }\textbf {\bibinfo {volume} {82}},\ \bibinfo {pages} {051404}
  (\bibinfo {year} {2010})}\BibitemShut {NoStop}%
\end{thebibliography}

%

\end{document}